\documentclass[12pt]{iopart}
\usepackage{iopams}
\usepackage{citesort}

\usepackage{CJK}                     
\usepackage{graphicx}                
\usepackage{bm}                      
\usepackage{mathptmx}                
\usepackage{xcolor}                  

\def\bea{\begin{eqnarray}} \def\eea{\end{eqnarray}}
\def\beq{\begin{equation}} \def\eeq{\end{equation}}
\def\bal#1\eal{\begin{align}#1\end{align}}
\def\bse#1\ese{\begin{subequations}#1\end{subequations}}
\def\non{\nonumber}

\def\text{\mathrm}
\def\be{\beta}
\def\om{\omega}
\def\eps{\varepsilon}
\def\la{\Lambda}
\def\tl{\tilde\Lambda}
\def\ms{M_\odot}
\def\mmax{M_\text{max}}

\begin{document}

\title
[Neutron star universal relations with microscopic equations of state]
{Neutron star universal relations with microscopic equations of state}

\begin{CJK*}{GBK}{}

\author{
J.-B. Wei$^1$,
A. Figura$^1$,
G. F. Burgio$^1$,
H. Chen$^2$
and H.-J. Schulze$^1$
}

\address{
$^1$ INFN Sezione di Catania and Dipartimento di Fisica,
Universit\'a di Catania, Via Santa Sofia 64, 95123 Catania, Italy}
\address{
$^2$ School of Mathematics and Physics,
China University of Geosciences, Lumo Road 388, 430074 Wuhan, P.R.~China}
\vspace{10pt}
\begin{indented}
\item[]September 2018
\end{indented}

\begin{abstract}
We calculate neutron star's moment of inertia and deformabilities
using various microscopic equations of state
for nuclear and hybrid star configurations.
We confirm several universal relations
between the various observables in these cases.
We focus in particular on the constraints for the neutron star radii
imposed by a determination of the average tidal deformability
of the binary neutron star system GW170817.
We find compatible radii between 12 and 13 kilometers
and identify the suitable equations of state.
\end{abstract}


\submitto{\jpg}



\end{CJK*}

\section{Introduction}

Neutron star (NS) observations allow us to probe the equation of state (EOS)
of nuclear matter \cite{eos,lattimer2016}
well beyond the densities available in terrestrial laboratories.
For example, observations of the NS mass--radius relation
\cite{lattimer2001,steiner2010,steiner2013,ozel16}
and the mass--moment-of-inertia relation \cite{schutz}
could ideally be used to infer the NS EOS within a
certain observational uncertainty.
While the masses of several NSs are known with good precision \cite{mass},
information on their radii is currently scarce and not very accurate
\cite{ozel16,gui2013,lat2014,steiner2016,steiner2018},
and direct measurements of the moment of inertia are so far not possible.
However, simultaneous measurement of both quantities for several objects
is required in order to constrain the EOS of NS matter
and allow conclusions regarding the composition of matter
under such extreme conditions.

The recent observation of gravitational waves emitted during the merger
of two corotating NSs \cite{merger} has opened the door
to new possibilities of obtaining information on the structural properties
of these objects, most prominently their masses and radii.
Binary NSs are, in fact, one of the most promising GW sources
\cite{faber2012,adligo}
for ground-based, second-generation detectors,
such as Advanced LIGO \cite{ligo} and Advanced VIRGO \cite{virgo}.

Comparing the observed GW signal with theoretical simulations,
some NS observables
have been identified as easily constrainable by a wave-form analysis.
Apart from the chirp mass \cite{pet63}
$M_c = {(M_1 M_2)^{3/5}}/{(M_1+M_2)}^{1/5}$
of the binary NS system,
in particular the tidal deformability \cite{hartle,flan}
$\lambda \equiv Q_{ij}/ E_{ij}$,
which measures the linear response of the quadrupole deformation $Q_{ij}$
to a (weak) external gravitational field $E_{ij}$,
could be well constrained by the new data.
In fact upper \cite{merger} and lower \cite{radice} limits
on the dimensionless quantity $\la\equiv\lambda/M^5$
have been deduced.
It is therefore of interest to examine these quantities
and their relations with other observables
in theoretical calculations of the EOS,
and numerous studies have already been performed in this sense
\cite{lattimer2001,steiner2010,steiner2013,radice,yagi13,yagi17}.

Also our current work follows this motivation
and we provide in this article the results obtained with several
`microscopic' EOSs, i.e.,
those based on many-body calculations employing fundamental free interactions.
Microscopic EOSs are
complementary to phenomenological EOSs with parameters fitted to
properties of nuclear matter and finite nuclei around saturation density,
and extrapolated to high density relevant for NSs \cite{2018Chap6}.
Relativistic-mean-field (RMF) models are usually used for this purpose
(see, e.g., the recent \cite{pascha,2018Malik,radice,tidal3,done} in the present context)
and offer by their nature more flexibility but less predictive power
than microscopic models.

We examine in particular several EOSs \cite{li08}
obtained within the Brueckner-Hartree-Fock (BHF) approach to nuclear matter
\cite{Jeu1976,baldo1999,bhf12},
and compare with the often-used results of the variational
calculation (APR) \cite{apr1998,morrales2002}
and the Dirac-BHF method (DBHF) \cite{dbhf1,dbhf2,dbhf3}.
Two phenomenological RMF EOSs are also included
for comparison:
LS220 \cite{ls} and SFHo \cite{sfh},
because they were recently used in an analysis of
and shown to be compatible with the GW170817/AT2017gfo event
\cite{shiba17,radice}.
Furthermore we also examine exotic variants containing hyperons
\cite{mmy1,mmy2,mmyy,chen11,chen12},
as well as hybrid stars obtained by allowing a Gibbs phase transition to
quark matter (QM) at high density.
We model the quark phase in the Dyson-Schwinger (DS) model
\cite{chen11,chen12,chen15,chen16,chen17a,chen17b},
and construct the phase transition to each of the nucleonic BHF models,
thus yielding several different hybrid EOSs detailed later.

We briefly comment that attempts to compute the nuclear EOS in
chiral perturbation theory \cite{epel,mach},
which has the theoretical advantage that two-body and three-body forces are
determined in a consistent way,
are still severely restricted by the perturbative low-density nature
of this approach.
Currently reliable results are limited to little more than
normal nuclear density,
at which point the regularization dependence becomes critical,
see, e.g., \cite{drischler,holt} for some recent illustrative results.

As stated before,
a stringent constraint on the EOS via the mass-radius relation
would be the measurement of both mass and radius of the same object
\cite{lattimer2001,steiner2010,steiner2013,ozel16}.
However, observations of NS radii are indirect and the determination
of the radius is affected by several uncertainties
(e.g., composition of the atmosphere, distance of the source, magnetic field;
see, e.g., \cite{sul2011,serv12,pot2014,fortin2015}.
Therefore precise estimations of NS radii are very difficult
because more model dependent than those of masses.
Currently, the most reliable constraints can be inferred from observations
of quiescent low-mass X-ray binaries (qLMXBs) in globular clusters
\cite{gui2013,2014Gui,2016Ozel},
because their atmospheres can be reliably modelled and their distances
can be accurately determined.
Constraints can also come from observations of type-I X-ray bursts
\cite{lattimer2001,steiner2010,steiner2013},
but this kind of analysis is still a matter of debate
\cite{mass,galdun2012,gui2013,guvoze2013,pout2014}.
Further information on radii could also be inferred from X-ray pulsation
in millisecond pulsars \cite{bog2016}.
Future high-precision telescopes and missions like
NICER \cite{nicer,nicer1,nicer2},
ATHENA+ \cite{athena,barcons2015}, and
SKA \cite{ska,watts2014}
are expected to improve our knowledge on the NS mass-radius relation.

This paper is organized as follows.
In Sec.~II we give a brief overview of the hadronic and hybrid EOSs we
are using.
In Sec.~III we specify some technical details regarding the computation
of moment of inertia and tidal deformability.
The numerical results obtained for those quantities along with the mass-radius
relation are presented in Sec.~IV for the different EOSs,
where we also investigate correlations with $M$ and $R$
and scrutinize various universality relations.
Sec.~V contains the conclusions.

\section{Equations of state}

In this section we will briefly discuss the EOSs used in this paper,
which are mainly microscopic EOSs based on many-body calculations.
For nuclear (hadronic) matter we resort to the BHF many-body theory
with realistic two-body and three-body nucleonic forces,
which has been extensively discussed in Refs.~\cite{Jeu1976,baldo1999,bhf12}.
In particular we examine several EOSs \cite{li08}
which are based on different nucleon-nucleon potentials,
the Argonne $V_{18}$ \cite{v18},
the Bonn~B (BOB) \cite{bonn1,bonn2},
and the Nijmegen~93 (N93) \cite{nij1,nij2},
and compatible three-nucleon forces as input.
More precisely,
the BOB and N93 are supplemented with microscopic TBF
employing the same meson-exchange parameters as the two-body potentials
\cite{glmm,zuo1,tbfnij},
whereas $V_{18}$ is combined either with a microscopic or a phenomenological TBF,
the latter consisting of an attractive term due to two-pion exchange
with excitation of an intermediate $\Delta$ resonance,
and a repulsive phenomenological central term
\cite{uix1,uix2,uix3,BCPM}.
They are labelled as V18 and UIX, respectively,
throughout the paper and in all figures.

The BHF theory has also been extended with the inclusion of hyperons,
which might appear in the core of a NS.
The hyperonic EOS in this theory turns out to be very soft,
and this results in too low NS maximum masses
\cite{mmy1,mmy2,chen11,chen12,mmyy},
$M<1.7\,\ms$ ($\ms\approx2\times10^{33}$g),
well below the current observational limit
of about two solar masses \cite{demorest2010,fonseca2016,heavy2}.
Nevertheless, such EOS could be realized in the so-called two-families scenario
in which the heaviest stars are interpreted as quark stars,
whereas the lighter and smaller stars are hadronic stars
\cite{drago1,drago2,drago3,drago4,pascha}.
We consider two BHF EOSs containing hyperons, which will be labelled as
BOB(NN+NY) \cite{mmy1,mmy2,chen11,chen12}
and V18(NN+NY+YY) \cite{mmyy}.

For completeness, we also compare with the often-used results of the
Dirac-BHF method (DBHF) \cite{dbhf1,dbhf2,dbhf3},
which employs the Bonn~A potential,
and the APR EOS based on the variational method \cite{apr1998,morrales2002}
and the $V_{18}$ potential.
The LS220 \cite{ls} and SFHo \cite{sfh} phenomenological
RMF EOSs are also used for comparison.

As far as the hybrid-star EOS is concerned,
it is widely known that the EOS for QM remains uncertain.
Whereas the microscopic theory of the nucleonic
EOS has reached a high degree of sophistication,
the QM EOS is poorly known at zero temperature and at the high baryonic density
appropriate for NSs,
because it is difficult to perform first-principle calculations of QM.
Therefore one can presently only resort to more or less phenomenological models
for describing QM,
such as the MIT bag model \cite{Chod}
or the Nambu--Jona-Lasinio model
\cite{Buballa2005,klahn2013,klahn2015}. 

The Dyson-Schwinger equations provide a nonperturbative continuum field
approach to QCD that can simultaneously address both confinement and dynamical
chiral symmetry breaking \cite{Roberts1994,Alkofer2000wg}.
In Refs.~\cite{chen11,chen12,chen15,chen16,chen17a,chen17b}
we developed a Dyson-Schwinger model (DSM) for deconfined QM
based on this formalism,
which was combined with the BHF approach for the hadronic phase
in order to model NSs.
We describe the quark phase in the DSM
with an interaction parameter $\alpha=1,2,3,4$
that models the quenching of the free quark-gluon vertex inside QM,
see Refs.~\cite{chen11,chen12,chen15,chen16,chen17a,chen17b} for details.
Increasing $\alpha$ leads to more stable QM in the DSM.
We then construct the Gibbs phase transition to each of the nucleonic models.
This yields 16 different EOSs
(BOB, V18, UIX, N93) $\otimes$ (DS1, DS2, DS3, DS4),
which differ essentially by their onset density of the QM phase
and the associated NS maximum mass,
both decreasing with increasing $\alpha$,
i.e., increasingly bound QM.

Properties of the various NS configurations constructed with the considered EOSs
are listed in Table~\ref{t:eos},
i.e., the value of the maximum mass, the corresponding radius,
the radius of the $1.4\,\ms$ configuration
and its tidal deformability $\la_{1.4}$,
which will be extensively discussed in Sect.~IV.
We mention that for the calculation of stellar structure we used the EOSs
of Refs.~\cite{fey,bps} for the outer crust
and \cite{NV} for the inner crust.
The choice of the crust model can influence the radius predictions
to a small extent,
of the order of $1\%$ for $R_{1.4}$ \cite{bhfls,2014Bur_Cent,fortin},
which is negligible for our purpose.
As far as the hybrid stars are concerned,
it is an important feature of the DSM that the hybrid star maximum mass
decreases with increasing QM fraction \cite{chen11,chen12},
and therefore
in the following we will consider only BOB+DS1, BOB+DS2, V18+DS1, and N93+DS1,
since only for those cases the static maximum mass is larger than 2 solar masses.

\def\myc#1{\multicolumn{1}{c}{$#1$}}
\begin{table}
\renewcommand{\arraystretch}{0.9}
\begin{center}
\caption{
\label{t:eos}
Properties of NSs listed according to the considered EOSs.
See text for details.}
\medskip
\begin{tabular}{@{}lrrrrll}
\hline
  EOS & \myc{\mmax[\ms]} & $R_{\mmax}$ [km] & $R_{1.4}$ [km] &
  $\phantom{-.}\la_{1.4}$ & Type & Ref. \\
\hline\\[-4mm]
  BOB      & 2.51 & 11.32 & 12.85 & 584 & nucleonic & \cite{li08}    \\
  BOB+DS1  & 2.30 & 12.13 & 12.85 & 584 & hybrid    & \cite{chen11}  \\
  BOB+DS2  & 2.02 & 11.95 & 12.85 & 584 & hybrid    & \cite{chen11}  \\
  BOB+DS3  & 1.79 & 11.72 & 12.75 & 539 & hybrid    & \cite{chen11}  \\
  BOB+DS4  & 1.60 & 11.38 & 12.12 & 346 & hybrid    & \cite{chen11}  \\
  V18      & 2.34 & 10.63 & 12.33 & 419 & nucleonic & \cite{li08}    \\
  V18+DS1  & 2.16 & 11.34 & 12.33 & 419 & hybrid    & \cite{chen11}  \\
  V18+DS2  & 1.93 & 11.15 & 12.33 & 419 & hybrid    & \cite{chen11}  \\
  V18+DS3  & 1.75 & 10.95 & 11.96 & 320 & hybrid    & \cite{chen11}  \\
  V18+DS4  & 1.61 & 10.74 & 11.36 & 215 & hybrid    & \cite{chen11}  \\
  N93      & 2.13 & 10.49 & 12.68 & 474 & nucleonic & \cite{li08}    \\
  N93+DS1  & 2.00 & 11.17 & 12.68 & 474 & hybrid    & \cite{chen11}  \\
  N93+DS2  & 1.80 & 10.76 & 12.64 & 459 & hybrid    & \cite{chen11}  \\
  N93+DS3  & 1.67 & 10.48 & 11.76 & 250 & hybrid    & \cite{chen11}  \\
  N93+DS4  & 1.58 & 10.31 & 11.05 & 162 & hybrid    & \cite{chen11}  \\
  UIX      & 2.04 & 10.02 & 12.03 & 340 & nucleonic & \cite{BCPM}    \\
  UIX+DS1  & 1.98 & 10.59 & 12.03 & 340 & hybrid    & \cite{chen11}  \\
  UIX+DS2  & 1.82 & 10.63 & 12.03 & 340 & hybrid    & \cite{chen11}  \\
  UIX+DS3  & 1.69 & 10.44 & 11.81 & 10  & hybrid    & \cite{chen11}  \\
  UIX+DS4  & 1.59 & 10.30 & 11.22 & 6   & hybrid    & \cite{chen11}  \\
  APR      & 2.20 &  9.92 & 11.59 & 274 & nucleonic & \cite{apr1998} \\
  DBHF     & 2.31 & 11.29 & 13.10 & 681 & nucleonic & \cite{dbhf3}   \\
  LS220    & 2.04 & 10.67 & 12.94 & 542 & nucleonic & \cite{ls}      \\
  SFHO     & 2.06 & 10.31 & 11.93 & 334 & nucleonic & \cite{sfh}     \\
  V18(N+Y) & 1.65 &  9.00 & 11.92 & 302 & hyperonic & \cite{mmyy}    \\
  BOB(N+Y) & 1.37 & 11.07 &  $-$  & $-$ & hyperonic & \cite{chen11}  \\
\hline
\end{tabular}
\end{center}
\end{table}

\section{Universal relations and global observables}
\label{s:form}


It is a major purpose of this work to confront the NS observables
and their relations
obtained with several microscopic EOSs listed in Table~\ref{t:eos}
to known universality relations.
Such universal (EOS-independent) relations between
the NS moment of inertia $I$,
the NS Love number $k_2$,
and the (spin-induced) NS quadrupole moment $Q$
(I-Love-Q relations)
are discussed in Refs.~\cite{yagi13,yagi17}.
Physically, the moment of inertia quantifies how fast a NS can spin
for a fixed angular momentum,
the quadrupole moment describes how much a NS is deformed away from sphericity,
and the Love number characterizes how easy it is to deform a NS.
These quantities can be computed
by numerically solving for the interior and exterior gravitational field
of a NS in a slow-rotation \cite{hartle}
and a small-tidal-deformation approximation
\cite{hinder2008,hinder2009,hinder2010}.
Although the moment of inertia is a first-order-in-spin quantity,
the quadrupole moment is generated by quadratic spin terms.
The tidal Love number \cite{tidal1,tidal2,tidal3}
is defined by the ratio between the
tidally-induced quadrupole moment and the tidal field due to a companion NS,
which can be calculated in a similar fashion.

One would expect that all of these quantities should depend quite sensitively
on the NS EOS;
instead they seem to satisfy almost universal relations
when plotted against each other in the proper way.
Possible explanations for this phenomenon are reviewed in
chapter~5 of \cite{yagi17},
but an ultimate formal proof is currently still missing.
Universal relations have various useful applications in astrophysics,
because the measurement of any member of the I-Love-Q trio automatically gives
the remaining two quantities without having to know the EOS.
The tidal Love number, for example,
has been constrained by Advanced LIGO and Advanced Virgo \cite{merger},
and by combining these results with the I-Love-Q relations,
one could equally constrain the moment of inertia and the quadrupole moment
of NSs in a binary system,
which would also be difficult to measure from GW observations.

In the following we briefly recall the formalism,
introducing the compactness parameter $\be=GM/Rc^2$,
with $G$ the gravitational constant and $c$ the speed of light.
Moreover we use geometrized units $G=c=1$.


The moment of inertia $I=J/\Omega$,
$J$ being the angular momentum,
and $\Omega \equiv 2\pi f$
the angular frequency measured by a distant observer (pulsar frequency),
the dimensionless ratio
\beq
 \frac{I}{MR^2} = \frac{1}{2\be} \frac{w_R}{3+w_R}
 \ , \quad
 w_R = \frac{r}{\om} \frac{d\om}{dr}\Big|_{r=R}
\eeq
[$w_R$ involving the metric function $\om$, Eq.~(\ref{e:metric}),
is obtained after integrating Eq.~(6)]
and the tidal deformability (quadrupole polarizability) $\lambda$,
$\la\equiv\lambda/M^5$,
or equivalently the tidal Love number $k_2$
\cite{hartle,hinder2008,hinder2009,hinder2010,tidal1,tidal2,tidal3},
\bea \fl
 k_2 = \frac{3}{2}\frac{\lambda}{R^5} = \frac{3}{2} \be^5 \la
\\ \fl\hskip5mm
 = \frac{8}{5} \frac{\be^5 z}{
 6\be(2-y_R) + 6\be^2(5y_R-8) + 4\be^3(13-11y_R) + 4\be^4(3y_R-2)
 + 8\be^5(1+y_R) + 3z\ln(1-2\be) } \:,
\non\\ \fl\hskip10mm
 z \equiv (1-2\be^2) [2-y_R+2\be(y_R-1)] \:,
 \label{eq:Love}
\eea
can be calculated in general relativity
together with the TOV equations
for pressure $p$ and enclosed mass $m$ of a static NS.
We follow the method outlined in
Refs.~\cite{tidal3,lattimer2016},
namely solve the system of four coupled first-order differential equations,
\bea
  {dp\over dr} &=& -{ m \eps \over r^2 }
  {  \left( 1 + {p/\eps} \right) \left( 1 + {4\pi r^3 p/m} \right)
  \over 1-2m/r } \:,
\\
  {dm \over dr} &=& 4 \pi r^2 \eps \:,
\\
  {dw \over dr} &=& \frac{4\pi r(\eps+p)(4+w)}{1 - 2m/r} - \frac{w(3+w)}{r} \:,
\\
  {dy \over dr} &=& -\frac{y^2}{r} - \frac{y-6}{r-2m} - rQ\:,
\non\\
  && Q \equiv 4\pi \frac{(5-y)\eps+(9+y)p+(\eps+p)/c_s^2}{1-2m/r}
  - \Big[ \frac{2(m+4\pi r^3 p)}{r(r-2m)} \Big]^2 \:,
\eea
with the EOS $\eps(p)$ as input and $c_s^2=d\eps/d\!p$.
The initial values are
\beq
 [p,m,w,y](r=0) = [p_c,0,0,2]
 \label{e:wr}
\eeq
and $w_R\equiv w(R)$, $y_R\equiv y(R)$.
In the case of an energy density discontinuity $\delta\eps$ in the EOS
(hydrid stars with Maxwell construction or pure quark stars without crust),
a separate finite contribution
\beq
 y_c = -3\delta\eps/\bar{\eps}
\eeq
has to be added to $y$ during the integration,
see Refs.~\cite{lattimer2001,steiner2010,steiner2013}.
The results have been compared with the output of the RNS code \cite{rns}
in the limit of vanishing rotation frequency,
and excellent agreement has been found.


In the case of an asymmetric binary system, $(M,R)_1+(M,R)_2$,
with mass asymmetry $q=M_2/M_1$,
and known chirp mass
\beq
 M_c = \frac{(M_1 M_2)^{3/5}}{(M_1+M_2)^{1/5}} \:,
\eeq
the effective deformability is given by
\beq
 \tl = \frac{16}{13}
 \frac{(1+12q)\la_1 + (q+12)\la_2}{(1+q)^5}
\label{e:lq}
\eeq
with
\beq
 \frac{[M_1,M_2]}{M_c} =
 \frac{297}{250} (1+q)^{1/5} [q^{-3/5},q^{2/5}] \:.
\eeq

The analysis of the GW170817 event \cite{merger} provided the data
$M_c/\ms=1.188{+0.004\atop-0.002}$
[corresponding to $M_1=M_2=1.365\,\ms$ for a symmetric binary system],
$q=M_2/M_1=0.7-1$
[corresponding to maximum asymmetry $(M_1,M_2)=(1.64,1.15)\,\ms$],
and $\tl<800$
from the phase-shift analysis of the observed signal.
The limit on $\tl$ was recently updated to
$70<\tl<720$ \cite{lv2018},
but we will keep using the original limit in this work.

\begin{figure}[t]
\vspace{-8mm}
\centerline
{\includegraphics[scale=0.40]{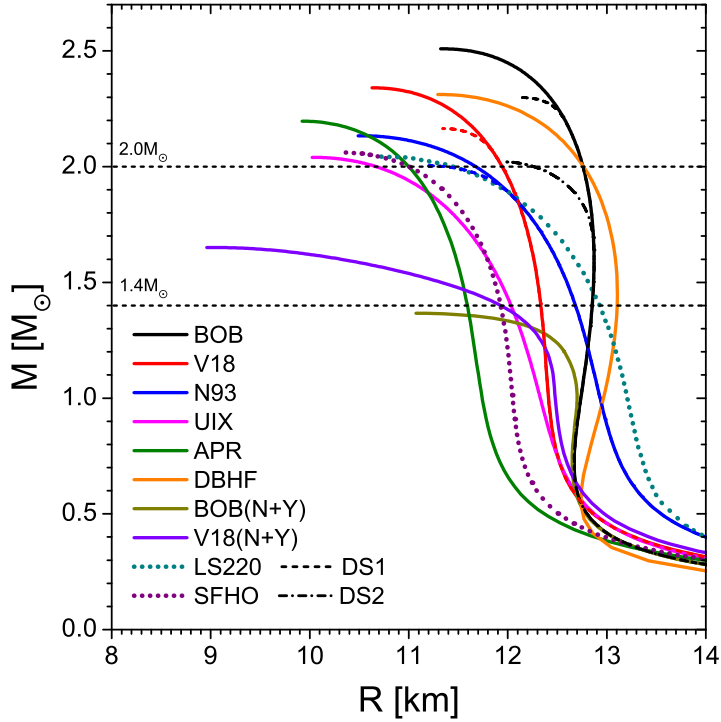}}
\vspace{-11mm}
\caption{
Mass-radius relations for different EOSs.
Solid (dotted) curves are plotted for microscopic (phenomenological) EOSs.
Dashed and dot-dashed lines indicate hybrid stars
in the DSM approach, see text.
}
\label{f:mr}
\end{figure}

\section{Results and discussion}

\subsection{Mass-radius relations}

Let us start by discussing the mass-radius relations of the different EOSs
we consider.
They are shown in Fig.~\ref{f:mr},
where results obtained with microscopic (phenomenological) EOSs are displayed
as solid (dotted) lines.
Moreover we consider the four EOSs for hybrid stars
with $\mmax>2\ms$ in Table~\ref{t:eos},
obtained by performing a Gibbs phase transition between the
BOB, V18, or N93 hadronic EOS and the DSM QM EOS
characterized by two different values of $\alpha=1,2$ (DS1, DS2).
They are displayed as dashed and dot-dashed lines, respectively.
We observe that most models give values of the maximum mass
larger than $2\,\ms$,
and therefore are compatible with current observational data
\cite{demorest2010,fonseca2016,heavy2}.
However, the two hyperonic EOSs do not fulfill the observational limit.
We nevertheless include them in our analysis in order to see whether they
reveal irregular features elsewhere.
Some recent analyses of the GW170817 event
indicate an upper limit of the maximum mass of about $2.2\,\ms$
\cite{shiba17,marga17,rezz18},
with which several of the microscopic EOSs would be compatible.

According to Fig.~\ref{f:mr}
(see also Table~\ref{t:eos}),
the predicted radii for a $M=1.4\,\ms$ NS span a range
$(11.6 \lesssim R_{1.4} \lesssim 13.2)\,$km.
Those values are in agreement with the ones reported in Ref.~\cite{annala17},
where an analysis of the results of GW170817 was performed
by using a general polytropic parametrization of the EOS
compatible with perturbative QCD at very high density.
In Ref.~\cite{annala17} it has been shown that the tidal deformability limit
of a $1.4\,\ms$ NS,
$\la_{1.4} < 800$,
as found in GW170817,
implies a radius $R_{1.4}<13.6\,$km.
Quite similar upper limits have been obtained in
\cite{most18,lim18,raithel18}.

Interpretation of the GW170817 event
also allowed to establish {\em lower} limits on the NS radius:
The condition of (meta)stability of the produced hypermassive star after merger
allowed to exclude very soft EOSs \cite{shiba17}
and to set thus a lower limit on the radius,
$R_{1.6} > 10.7\,$km \cite{baus17},
confirmed by similar recent analyses \cite{most18,lim18} in which
$R_{1.4}>(11.5-12)\,$km.
An even higher lower limit $R_{1.4}>12.55\,$km \cite{fatto17}
has been deduced from the measurement of the neutron skin of $\rm^{208}Pb$
in the PREX experiment \cite{abra12}.
Simulations with several different EOSs set also a lower limit
on the effective deformability Eq.~(\ref{e:lq}),
$\tl>400$ \cite{radice},
related to the black hole formation time
and the accretion disk mass of material left out of the black hole.
The latter was constrained from optical/infrared observations
of the remnant AT2017gfo \cite{knova1,knova2,knova3,knova4,knova5}.
Small values of $\tl$ and therefore small values of $R$ imply
very fast black hole formation and little material left in the disk,
which is incompatible with observation.
A correlated lower limit $R_{1.4} \gtrsim 12\,$km is obtained in this way.

On the other hand,
smaller radii than these lower limits were deduced
from observations of thermal emission from accreting NSs in quiescent LMXBs.
In fact, by analyzing their X-ray spectra,
the observations seem to suggest for stars of mass about $(1.4-1.5)\,\ms$
a radius in the range $(9.9-11.2)\,$km \cite{ozel16}.
Those results have been criticized in
\cite{lattimer2001,steiner2010,steiner2013}:
in particular the estimates of the radii are affected by the uncertainties
on the composition of the atmosphere.
If the atmosphere contains He,
significantly larger radii are extracted.
More recently \cite{steiner2018}
it was shown that when allowing for the
occurrence of a first-order phase transition in dense matter (Model C),
$R_{1.4}$ is smaller than $12\,$km to $95\%$ confidence.
However, $R_{1.4}$ could be larger if NSs have uneven temperature
distributions.
Clearly, no firm conclusions can yet be reached and we need to wait
for new data such as the ones collected by the NICER mission,
in order to obtain independent and precise information on NS radii.

We remark that this clash between large radii from GW170817
and small radii from quiescent LMXBs (if confirmed)
could be resolved in the two-families or twin-star scenarios,
in which small and big stars of the same mass
could coexist as hadronic and QM stars
\cite{drago1,drago2,drago3,drago4,pascha}.

\begin{figure*}[t]
\vspace{-15mm}
\centerline{
\includegraphics[scale=0.65]{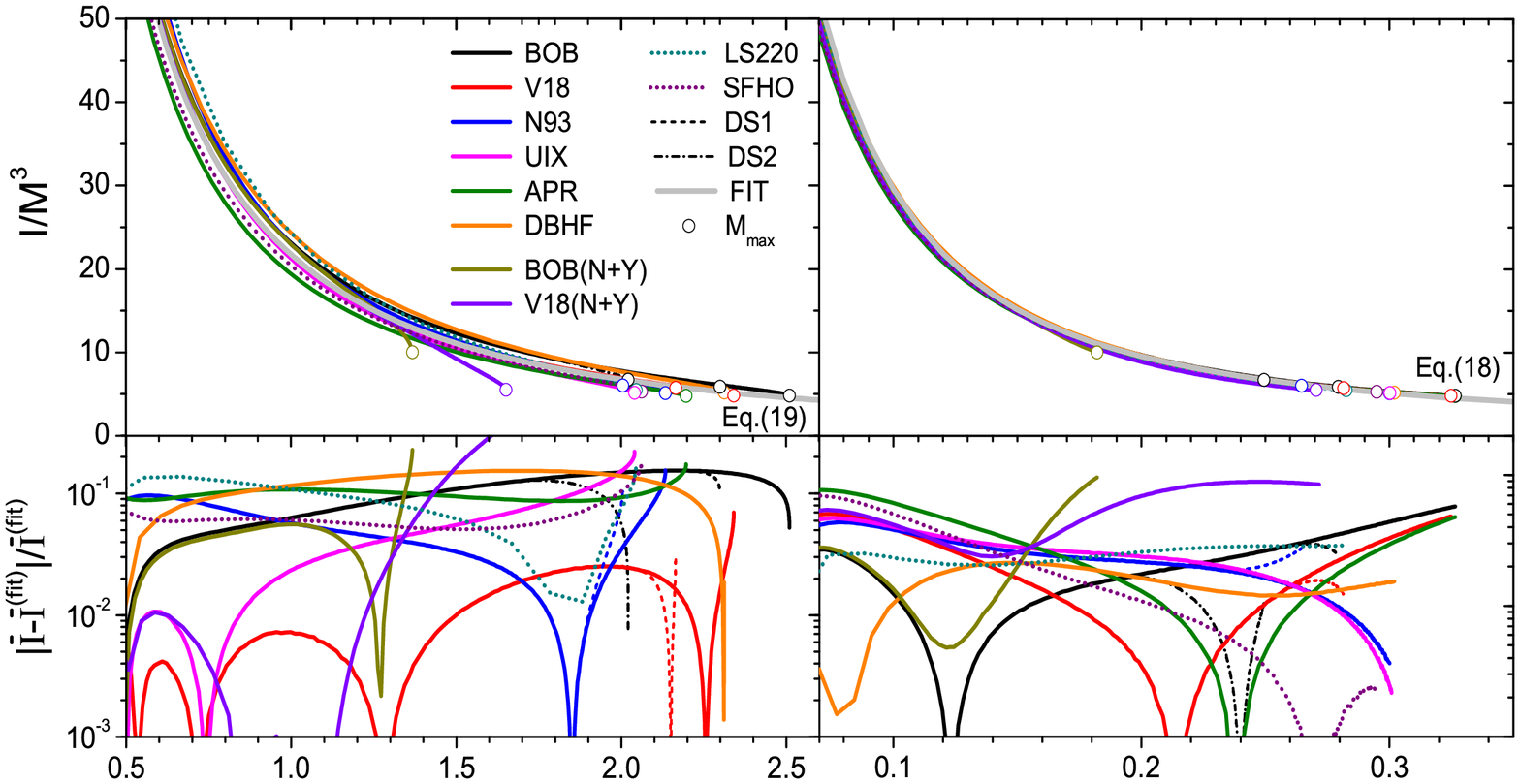}}
\label{f:i}
\end{figure*}

\begin{figure*}[t]
\vspace{-28mm}
\centerline{
\includegraphics[scale=0.65]{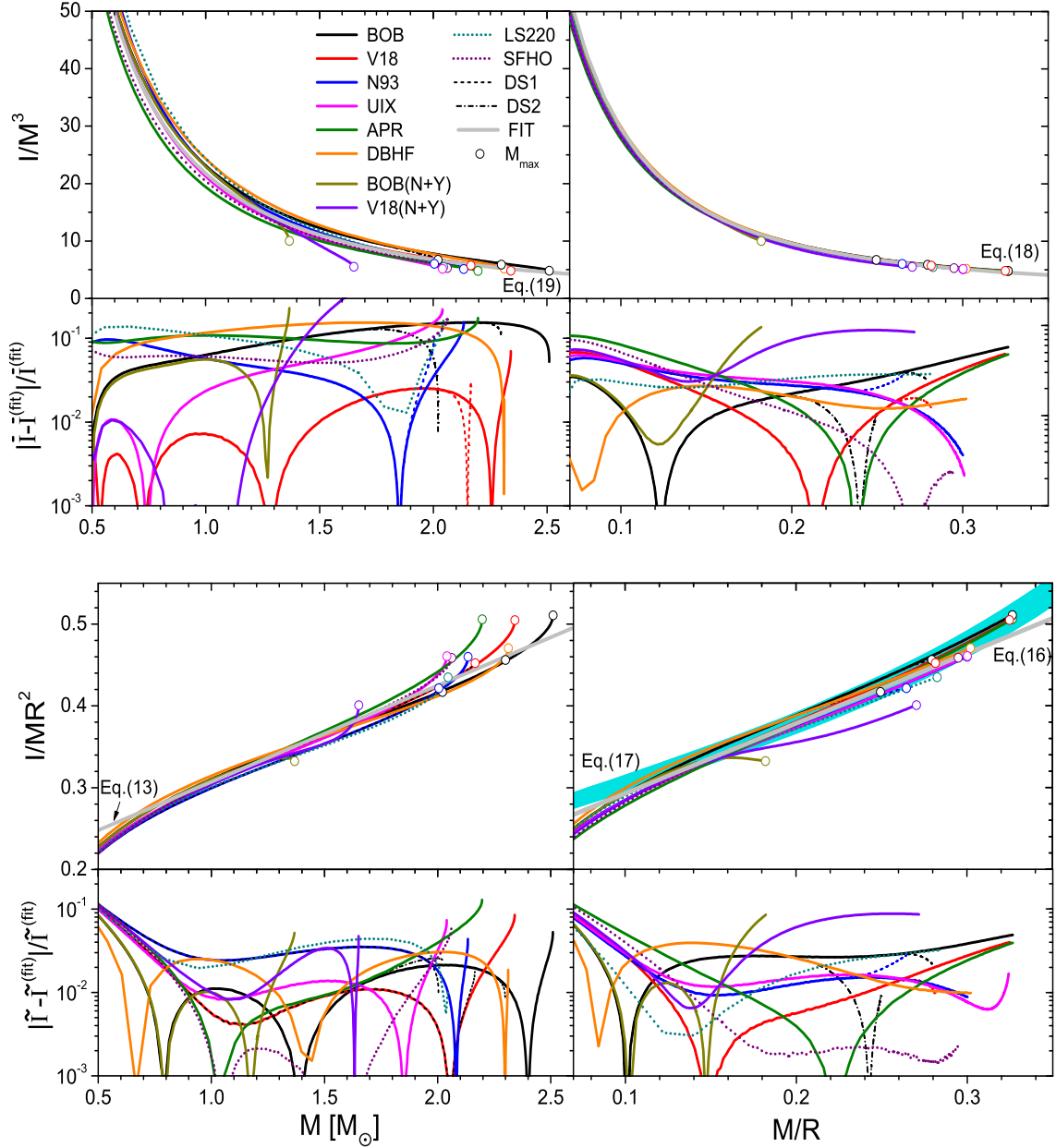}}
\vspace{-20mm}
\caption{
$I/MR^2$ (lower panels)
and
$I/M^3$ (upper panels)
vs.~$M$ (left panels) and $M/R$ (right panels)
for the 10+4 different nucleonic+hybrid EOSs shown in Fig.~\ref{f:mr}.
Configurations of $M=\mmax$ are indicated by markers.
The grey curves indicate the fits according to Eqs.~(\ref{e:13}),(\ref{e:16}),(\ref{e:breu}),(\ref{e:19}).
The fit Eq.~(\ref{e:fitmb}) taken
from Refs.~\cite{lattimer2001,steiner2010,steiner2013}
is shown for comparison as a blue band, see text.
In each panel, the upper part shows the results for the different EOSs,
and the lower part the fractional deviations from the grey fit curves.
}
\label{f:ibis}
\end{figure*}

\subsection{The moment of inertia $I$}

Taking for granted the validity of a universality relation,
the moment of inertia $I$ of a NS can be expressed as a function
of the NS mass and radius,
and therefore the radius could be determined if the mass and the
moment of inertia of the NS is known \cite{schutz,wor}.
Dimensionally, the moment of inertia is proportional to the star's mass
times its radius squared,
so a measurement of the moment of inertia to a given accuracy
provides approximately twice that accuracy for a radius identification.
As already stated in Ref.~\cite{schutz},
estimating a NS moment of inertia from timing observations
of a binary radio pulsar has significant implications for constraining the EOS.
In some respect, a measurement of the moment of inertia could be more useful
than a radius measurement of the same accuracy.
However,
the moment of inertia of a rotating NS has not yet been measured directly.

In Fig.~\ref{f:ibis} we show the moment of inertia
normalized in two different ways,
$I/MR^2$ (lower panels) and
$I/M^3$ (upper panels)
vs.~the gravitational mass $M$ (left panels)
and the compactness $\beta$ (right panels),
for the same EOSs as shown in Fig.~\ref{f:mr}.

We observe in general that all curves lie within ``universality bands," i.e.,
they are nearly EOS independent.
For example, the universal relation in the $I/MR^2$ vs.~$M$ plot,
indicated by a grey curve,
is quantified as a simple linear fit,
valid in the interval $1<M/\ms<2$,
\beq
 \frac{I}{MR^2} 
 \approx  0.189 + 0.118 \frac{M}{\ms} \pm 0.016 \:.
\label{e:13}
\eeq
For fixed (measured) $M$ and $I$ and unknown $R$ we have then
\beq
 \delta(f \equiv \frac{I}{MR^2}) = - \frac{2I}{MR^3} \delta R
 = -2f \frac{\delta R}{R}
\eeq
and thus the universality band determines $R$ with accuracy
\beq
 \frac{\delta R}{R} = \frac{1}{2}\frac{\delta f}{f}
 \approx \frac{0.016}{0.4} \lesssim 4\% \:.
\eeq

For the $I/MR^2$ vs.~$M/R$ plot,
the equivalent fit for our chosen set of microscopic EOSs reads
(grey curve)
\beq
 \frac{I}{MR^2} \equiv 0.207 + 0.857\beta \pm 0.011 \:,
\label{e:16}
\eeq
to be compared with the one reported in
Ref.~\cite{lattimer2001,steiner2010,steiner2013},
obtained with a larger set of EOSs,
displayed as a blue band,
\beq
 \frac{I}{MR^2} \approx
 (0.237 \pm 0.008) (1 + 2.844\beta + 18.91\beta^4)
\label{e:fitmb}
\eeq
It may be noted that the fit vs.~$M$ fails mainly for large masses $M>2\,\ms$,
whereas the fit vs.~$\be$ fails for hyperonic stars with low $\mmax$.
The latter feature is caused by the small radius of the
maximum mass configuration for the hyperonic EOSs, see Fig.~\ref{f:mr},
which leads to an `abnormaly' large value of $\beta=M/R$ close to the
(small) maximum mass.
Turning the universality argument around,
future simultaneous measurement of $M,R,I$
could therefore provide evidence for the presence of hyperons in a NS,
at least close to the maximum mass configuration.

The upper panels of Fig.~\ref{f:ibis} show the results obtained for the
quantity $I/M^3$ as advocated in \cite{breu},
together with the fits
\beq
 \frac{I}{M^3} \equiv 0.8134 \,\beta^{-1} + 0.2101\, \beta^{-2}+ 0.003175\, \beta^{-3} - 0.0002717\, \beta^{-4}
\label{e:breu}
\eeq
obtained in \cite{breu} and
\beq
 \frac{I}{M^3} \equiv 1.0334\, M^{-1} + 30.7271 \,M^{-2}- 12.8839 \, M^{-3} + 2.8841 \,M^{-4}
\label{e:19}
\eeq
obtained by us.
Summarizing, in all four panels (except the case of $I/M^3$ vs. $M$ plot), the deviations of the individual EOSs from the universal fits
are of the order of a few percent, largest with the hyperonic EOSs.

\subsection{The tidal deformability $\la$}

\begin{figure}[t]
\vspace{-11mm}
\centerline{\hskip7mm\includegraphics[scale=0.49]{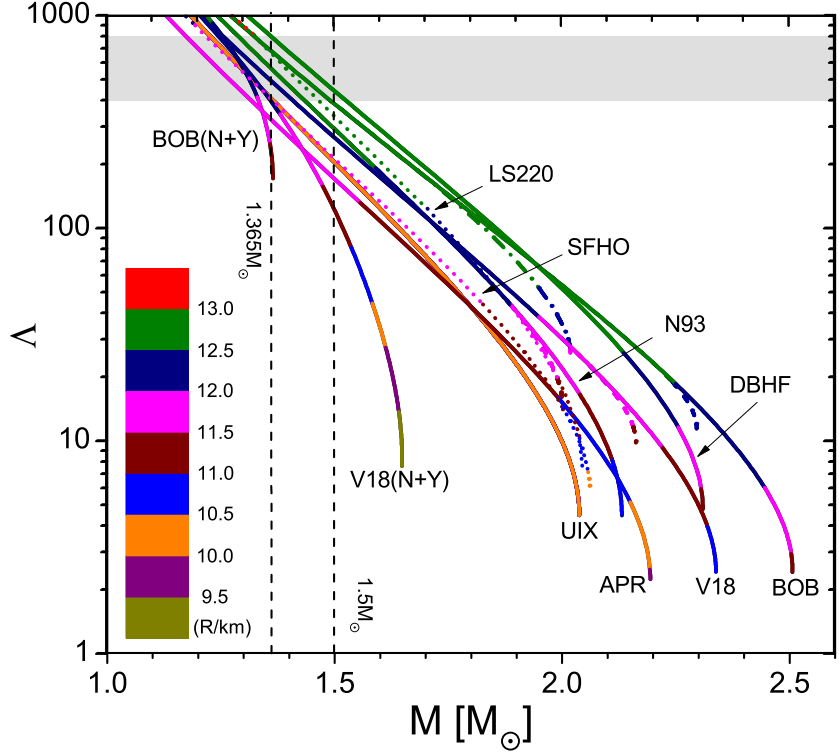}}
\vspace{-6mm}
\caption{
Correlations between $M$, $R$, and $\la$ for a single NS with different EOSs,
see Fig.~\ref{f:mr}.
Dashed and dot-dashed curves display hybrid stars with DS1 and DS2, respectively.
The shaded area is constrained by the interpretation of the GW170817 event
as a symmetric NS merger.
}
\label{f:lm}
\end{figure}

We now turn to discuss a further important global observable,
the tidal deformability $\la$ of a single NS,
see Eq.~(\ref{eq:Love}).
Fig.~\ref{f:lm} shows $\la$
as a function of $M$ and $R$ for the various EOSs considered in this paper.
The information on the radius $R$ is encoded in the colored segments
of the curves.
The grey band represents the observational limits
derived in Refs.~\cite{merger,radice}
mentioned before,
$400<\la<800$,
deduced from a multimessenger analysis of the GW170817 event
on the amplitude of tidal effects during the binary inspiral,
combined with an analysis of the UV/optical/infrared counterpart
with kilonova models.
The black dashed line at fixed mass $M=1.365\,\ms$ indicates the masses
of each NS for a symmetric binary system in GW170817,
whereas the line located at $1.5\,\ms$ represents the constraint
derived in Ref.~\cite{annala17}.

One notes that the conditions $M=1.365\,\ms$ and $400<\la<800$
imply $12\,$km$\,\lesssim R \lesssim 13\,$km,
with the compatible EOSs V18(N+Y), UIX, V18, N93, BOB, DBHF
in order of increasing radius
(see also Table~\ref{t:eos}).
Also the phenomenological EOS labelled LS220 fullfills the constraint,
as well as trivially the hybrid stars constructed with DS1 and DS2 EOSs,
which at $M=1.365\,\ms$ are still purely nucleonic.
On the other hand, APR, BOB(N+Y), and SFHO (marginally)
do not fulfill the $\la>400$ constraint.

\begin{figure*}[t]
\vspace{-9mm}
\centerline{\includegraphics[scale=0.67]{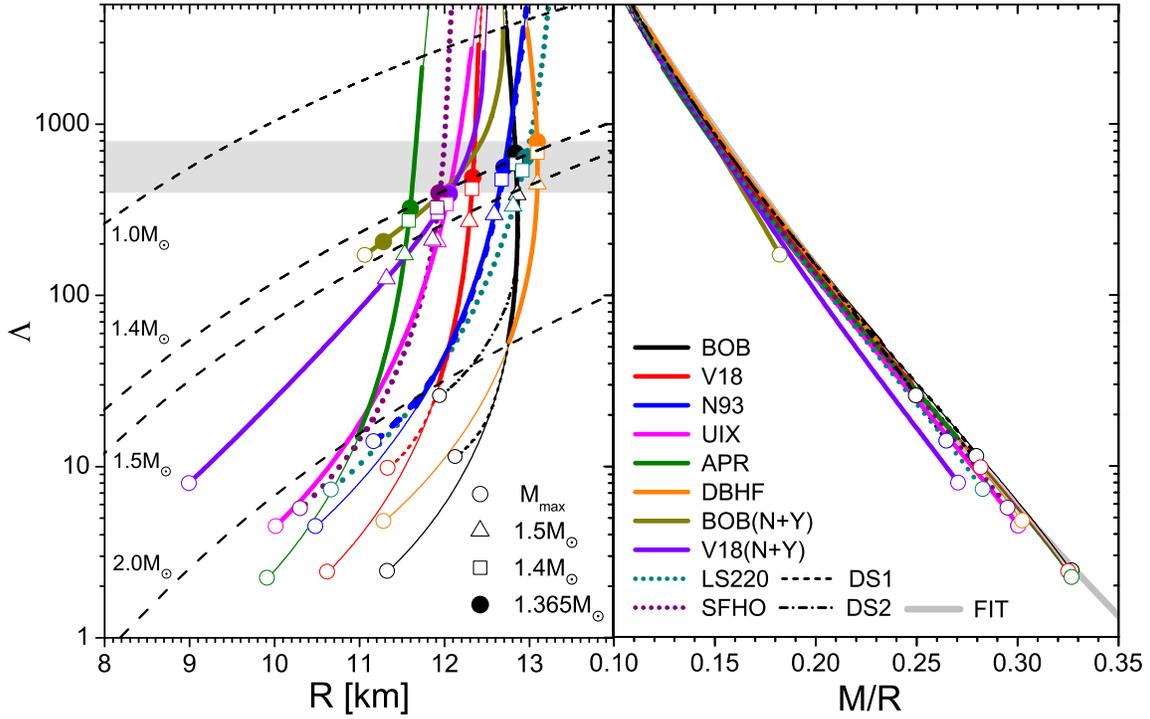}}
\vspace{-8mm}
\caption{
Correlations between $\la,M,R$
for a single NS with different EOSs.
The shaded area is constrained by the interpretation of the GW170817 event
as a symmetric NS merger.
The dashed (left panel)
or grey (right panel)
curves show the predictions according to the universal fit
Eq.~(\ref{e:univ}).
}
\label{f:lr}
\end{figure*}

The same kind of information can be derived also by displaying
the tidal deformability $\la$ of a single NS as a function of the radius $R$
and the compactness $\beta$ for the considered EOS,
and this is shown in Fig.~\ref{f:lr}.
As before, the grey band represents the constraints discussed above,
and the different curves are obtained for the different EOSs.
The curves have a bold width in the interval $1<M/\ms<2$.
The full circles represent the ($\la$,$R$) configurations at $M=1.365\,\ms$,
whereas the open squares and triangles indicate
$M=1.4\,\ms$ and $M=1.5\,\ms$ configurations, respectively.
The endpoints, displayed as open circles,
represent the maximum mass for that chosen EOS.

\begin{figure}[t]
\vspace{-6mm}
\hspace{-5mm}\includegraphics[scale=0.64]{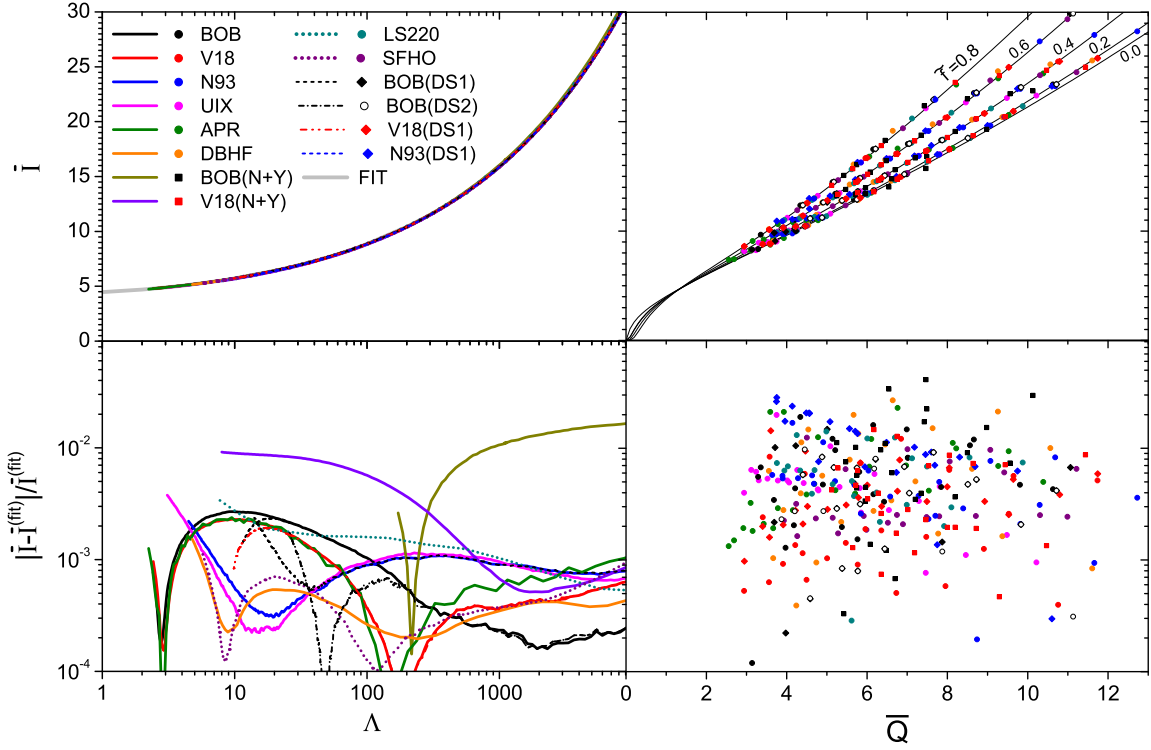}
\vspace{-12mm}
\caption{
Relation between dimensionless quantitities $\bar{I}=I/M^3$
and $\la$
(left plot)
or $\bar Q= -{QM}/{(I\Omega)^2}$
(right plot)
in comparison with the fits Eq.~(\ref{e:il}) or
Eqs.~(\ref{e:iq0},\ref{e:iq}) (curves), respectively.
The colored markers in the right plot
show results obtained with different EOSs
at fixed rotation parameter
$\tilde{f} \equiv 20 R f = 0.0,0.2,0.4,0.6,0.8$.
In the bottom panels,
relative errors between the fitting curves and numerical results
are displayed.
}
\label{f:il}
\end{figure}

A universal relation of the individual tidal deformabilities of NSs
as function of the stellar compactness was introduced in Ref.~\cite{yagi13},
and in Ref.~\cite{yagi17} the following fit was proposed
\beq
 \be = 0.36 - 0.0355 \ln\la + 0.000705 (\ln\la)^2 \:,
\label{e:univ}
\eeq
\noindent
which holds to within 6.5$\%$ for a large set of NS EOSs \cite{yagi17}.
The dashed black lines in the left panel and the solid grey line
in the right panel represent the results of Eq.~(\ref{e:univ}),
and particularly in the right panel one can observe
that the fit works well
(slighly overestimating the values of $\la$)
also for our set of microscopic EOSs,
except for the very soft ones including hyperons.

For completeness we confirm also the
extremely tight universal relation between the dimensionless quantities
$\bar{I} \equiv I/M^3$   
and $\la$   
\cite{yagi17},
namely
\bea
 \ln(I/M^3) &=& 1.496 + 0.05951 \ln\la + 0.02238 (\ln\la)^2
\nonumber\\&& 
  - 6.953\times10^{-4} (\ln\la)^3 + 8.345\times10^{-6} (\ln\la)^4
\:.\hskip3mm
\label{e:il}
\eea
The comparison with our EOSs is shown in Fig.~\ref{f:il} (left plot)
and confirms the validity of the fit (grey curve).
The bottom of this figure shows the fractional errors between the
fitting function and the numerical results,
and they amount to less than $1\%$,
being a bit larger for EOSs including hyperons.

\subsection{The quadrupole moment $Q$}

A further universal, i.e., EOS-independent, relation
involves the moment of inertia $I$ and the spin-induced quadrupole moment $Q$,
as proposed in Refs.~\cite{yagi13,yagi17}.
This relation was derived in the slow-rotation and
small-tidal-deformation approximations,
leaving open the question of its general validity,
faced in some successive papers \cite{done,chakra}.
Here, we analyze the validity of this universal relation
for our set of microscopic EOSs.

For analyzing the field equations of the rotating NS and compute $Q$,
we use the RNS code \cite{rns},
which assumes steady rotation and axisymmetric structure.
Therefore the space-time metric can be expressed as \cite{cipolletta}
\bea
 ds^2 &=& -e^{\gamma+\rho} dt^2 + e^{2\beta} (dr^2+r^2d\theta^2)
 + e^{\gamma-\rho} r^2\sin^2\!\theta (d\phi-\om dt)^2 \:,
\label{e:metric}
\eea
where the potentials $\gamma,\rho,\beta$,$\om$ are functions of
$r$ and $\theta$ only.
The quadrupole moment calculated using RNS, $Q^\text{RNS}$,
has to be corrected \cite{ryan,geroch,hansen}
and is given by \cite{pappas,cipolletta}
\beq
 Q = Q^\text{RNS} -\frac{4}{3}\left( \frac{1}{4} +b_0 \right) M^3 \:,
\eeq
where $M$ is the mass of the star,
and the parameter $b_0$ is given by
\bea
 b_0 &=& -\frac{16\sqrt{2\pi}r_\text{eq}^4}{M^2}
 \int_0^{\frac{1}{2}} \frac{s^3ds}{(1-s)^5}
 \int^1_0 d\mu
 \sqrt{1-\mu^2} \,P(s,\mu) e^{\gamma+2\beta} T_0^\frac{1}{2}(\mu) \:.
\eea
Here
$r_\text{eq}$ is the value of the coordinate radius at the equator,
$s=r/(r+r_\text{eq})$ is a compacted radial coordinate,
$\mu=\cos(\theta)$,
$P(s,\mu)$ is the pressure, and
$T_0^{\frac{1}{2}}(\mu)=\sqrt{2/\pi}C_0(\mu)$
with $C_0$ the 0th-order Gegenbauer polynomial.

In order to investigate universal relations,
the following dimensionless quantities
were introduced \cite{chakra}
\bea
 \bar{I} \equiv \frac{I}{M^3}
\:, \quad
 \bar{Q} \equiv -\frac{QM}{(I\Omega)^2} 
\:.
\eea
In Fig.~\ref{f:il} (right plot) we display $\bar I$ vs.~$\bar Q$
for the various EOSs and
for different normalized rotational frequencies
$\tilde{f} \equiv 20 R f = 0.0,0.2,0.4,0.6,0.8$.
(The normalization is such that $f=1$~kHz corresponds
to $\tilde {f} \approx 1$ for $R=15\,$km).
The $\tilde f=0.0$ curve is the one of Refs.~\cite{yagi13,yagi17},
obtained in the limit of small frequency,
\bea
 \fl\hskip4mm
 \ln{\bar I} &=& 1.393 + 0.5471 \ln \bar Q + 0.03028 (\ln \bar Q)^2
 +\, 0.01926  (\ln \bar Q)^3 + 4.434 \times10^{-4} (\ln \bar Q)^4
\:.\hskip6mm
\label{e:iq0}
\eea
The other curves represent the refined fits given in Ref.~\cite{chakra},
where an explicit dependence of the above coefficients
on $\tilde f$,
or alternatively on the parameter $a \equiv I \Omega/M^2$, 
was introduced for the range
$0.2 < \tilde f < 1.2$,
$0.1 < a < 0.6$,
$1.5 < \bar Q < 15$:
\beq
 \ln{\bar I}
 \approx \sum_{i,j=0,4}  \mathcal A_{ij} a^i (\ln\bar Q)^j
 \approx \sum_{i,j=0,4}  \mathcal B_{ij} \tilde f^i (\ln\bar Q)^j \:,
\label{e:iq}
\eeq
with the parameters $ \mathcal A_{ij}$ and $ \mathcal B_{ij}$
to be found in Ref.~\cite{chakra}.

One observes that also for our set of EOSs the fit works very well,
the relative deviations remaining below three percent in most cases,
as shown in the lower panel of Fig.~\ref{f:il} (right).
As motivated in Ref.~\cite{chakra},
this allows in principle the determination of the NS radius
appearing in the definition of $\tilde f=20Rf$,
if the correlated quantities $\bar I,\bar Q$ are sufficiently well known.
We notice, however, that the spin-induced quadrupole moment  $Q$ is largely degenerate with the
mass ratio and the spins, and this makes it very difficult to measure independently
\cite{krish}.

\section{Conclusions}
\label{s:end}

We have confirmed the validity of several universal relations
among the moment of inertia $I$,
the tidal deformability $\la$,
and the quadrupole moment $Q$,
as proposed some years ago \cite{yagi13,yagi17}.
In particular,
we have examined several microscopic EOSs
obtained within the BHF approach to nuclear matter,
along with the well-known variational APR and the DBHF EOS,
and compared with two phenomenological RMF EOSs, LS220 and SFHo.
We have also analyzed the BHF EOS for stellar matter containing hyperons,
as well as hybrid stars with quark matter at high density,
modeled in the Dyson-Schwinger theoretical framework.
The strongest deviations from universality are exhibited by the hyperonic EOSs,
with the eventual possibility to identify their presence in NSs in this way.

We have demonstrated that the microscopic equations of state
derived some time ago in the BHF formalism and
based on meson-exchange nucleon-nucleon potentials
and consistent microscopic three-body forces,
are fully compatible with new constraints imposed by interpretation of the
first observed neutron-star merger event GW170817.
In particular, they respect the lower $2\ms$ limit of the NS maximum mass
and feature typical radii between 12 and 13 km,
constrained by tightly correlated values of the tidal deformability $\la$.
The same holds true also for the relativistic DBHF EOS.

We stress that all results presented here were obtained by assuming that
Einstein's general relativity (GR) is the correct theory of gravity.
However, NS are also unique probes of strong-field gravitational physics,
and therefore extensions of GR have to be taken into account.
For any given EOS, theories that modify the strong-field dynamics of GR
generally predict static properties, e.g.,
mass, radius, and moment of inertia,
that are different from those in Einstein's theory.
The rich literature available on NS treated in modified theories of gravity
reveals a high degree of degeneracy in the main properties of relativistic stars
\cite{Berti15},
and this highlights the potential of future GW measurements to
determine the behavior of gravity in the strong-field regime.

The detection of gravitational waves by the LIGO/Virgo collaboration in 2015,
and the successive binary neutron star merger event GW170817,
opened a new astronomical eye to the Universe,
and NSs play a key role in this respect,
having the potential of being extremely prolific gravitational wave emitters
in terms of expected detection rates.
Therefore we are looking forward to more refined constraints
to be obtained soon from further merger events and new facilities.

\ack

We acknowledge useful discussions with A.~Drago and G.~Pagliara.
H. Chen acknowledges financial support from NSFC (11475149, U1738130),
and J.-B. Wei acknowledges the China Scholarship Council
(CSC File No.~201706410092) for financial support.
Partial support comes also from ``PHAROS," COST Action CA16214.

\section*{References}

\bibliographystyle{iopart-num}
\bibliography{miclam}

\end{document}